\documentclass{emulateapj}
\usepackage{times}

\slugcomment{ApJL, accepted}

\shorttitle{X-ray luminous AGN in massive galaxy clusters}
\shortauthors{Ruderman and Ebeling}

\begin{document}

\title{The Origin of the Spatial Distribution of X-ray luminous AGN in
Massive Galaxy Clusters}


\author{Joshua T.\ Ruderman\altaffilmark{1} and Harald Ebeling}

\affil{Institute for Astronomy, University of Hawaii, 2680 Woodlawn
Drive, Honolulu, HI 96822} 

\altaffiltext{1}{also Departments of Physics and Mathematics, Stanford
University, Stanford, CA 94305}

\begin{abstract}
We study the spatial distribution of a 95\% complete sample of 508
X-ray point sources (XPS) detected in the 0.5--2.0 keV band in Chandra
ACIS-I observations of 51 massive galaxy clusters found in the MACS
survey. Covering the redshift range $z=0.3-0.7$, our cluster sample is
statistically complete and comprises all MACS clusters with X-ray
luminosities in excess of $4.5\times 10^{44}$ erg s$^{-1}$ (0.1--2.4
keV, $h_0=0.7$, $\Lambda$CDM). Also studied are 20 control fields that
do not contain clusters. We find the XPS surface density, computed in
the cluster restframe, to exhibit a pronounced excess within 3.5 Mpc
of the cluster centers. The excess, believed to be caused by AGN in
the cluster, is significant at the $8.0\sigma$ confidence level
compared to the XPS density observed at the field edges.  No
significant central excess is found in the control fields. To
investigate the physical origin of the AGN excess, we study the radial
AGN density profile for a subset of 24 virialized clusters. We find a
pronounced central spike ($r<0.5$ Mpc), followed by a depletion region
at about 1.5 Mpc, and a broad secondary excess centered at
approximately the virial radius of the host clusters ($\approx 2.5$
Mpc). We present evidence that the central AGN excess reflects
increased nuclear activity triggered by close encounters between
infalling galaxies and the giant cD-type elliptical occupying the very
cluster center. By contrast, the secondary excess at the cluster-field
interface is likely due to black holes being fueled by galaxy mergers.
In-depth spectroscopic and photometric follow-up observations of the
optical counterparts of the XPS in a subset of our sample are being
conducted to confirm this picture.
\end{abstract}

\keywords{galaxies: active --- galaxies: clusters: general --- galaxies: evolution --- galaxies: interactions --- X-rays: galaxies --- X-rays: galaxies: clusters}


\section{Introduction}

The abundance and properties of AGN in galaxy clusters are important
diagnostics for studies of cluster formation and galaxy evolution, but
are difficult to measure as a large number of homeogenously selected
systems is required for a statistically robust result. Massive galaxy
clusters should be preferred targets for this kind of research as they
constitute the largest reservoirs of galaxies. On the other hand, the
galaxy population specifically of evolved clusters is overwhelmingly
dominated by elliptical galaxies which, at least in the local
universe, are gas-poor and thus do not typically exhibit nuclear
activity. The situation is complicated further by the fact that, until
recently, observational studies of the AGN distribution in clusters
and the field have based AGN identifications almost exclusively on
characteristic emission lines in the optical part of the galaxy
spectrum, making them insensitive to a hypothetical population of
extremely obscured AGN.

In the past few years, several studies have taken a complementary
approach by using X-ray observations, specifically the surface density
of X-ray point sources (XPS) around galaxy clusters, to quantify the
AGN content of clusters relative to that of the field. Evidence of an
XPS excess has been presented for several clusters over a wide range
of redshifts and X-ray luminosities
\citep{henry,cappi,molnar,sun,cappelluti}. If the point sources
detected in these fields are indeed at the cluster redshift, their
X-ray luminosities suggest that the excess is caused by AGN in the
cluster and can thus be used to map the cluster's AGN distribution.
Optical follow-up observations performed in the field of A2104 at
$z=0.154$ \citep{martini} confirm this picture and reveal a possibly
considerable fraction of these galaxies to lack the spectroscopic
characteristics of AGN, suggesting that clusters may contain a large
subset of optically obscured AGN.

Using Chandra ACIS-I data we here present the results of the first
systematic study of the XPS content of a large, representative sample
of massive galaxy clusters, to test the findings of earlier studies
obtained for individual clusters. The large number of XPS detections
obtained in this work allows us to construct the first resolved radial
profile of XPS in the cluster rest frame, and to characterize the
spatial XPS distribution as well as its likely physical origin.
Throughout we assume a $\Lambda$CDM cosmology with $h_{\rm 0}$=0.7,
$\Lambda$=0.7, and $\Omega$=0.3.\\

\section{Field selection}

\begin{figure}
\epsscale{1.1}
\plotone{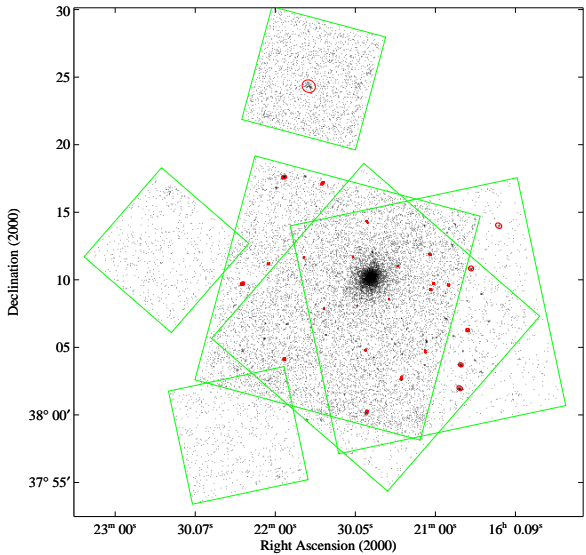}
\caption{\label{fig1} Merged dataset constructed from the three ACIS
observations of the relaxed cluster MACS\,J1621.3+3810 at $z=0.46$
\citep{edge}. Red ellipses mark the point sources detected with {\it
Celldetect} using a $3\sigma$ detection threshold. The green
rectangles show the orientation of the front-illuminated chips for the
three individual exposures.}
\end{figure}

We analyze Chandra ACIS data (front-illuminated chips only) for all 51
MACS clusters (MACS = MAssive Cluster Survey) \citep{ebeling} at $z=
0.3-0.7$ that have been observed before Jan 15, 2005 and feature X-ray
luminosities in excess of $4.5\times 10^{44}$ erg s$^{-1}$ (0.1--2.4
keV). The total geometric area covered by these observations is 5.73
deg$^2$. An identical analysis is performed for ACIS-I data of 20
control fields, i.e. observations that did not target galaxy clusters,
covering a total of 1.98 deg$^2$. On-axis exposure times range from 10
ks to 87 ks.  We reduce the raw data using CIAO software version 3.02,
removing systematic instrumental effects and background flares. Merged
datasets are created for the nine clusters observed twice and three
clusters observed three times (see Fig.~\ref{fig1} for an example). To
account for exposure-time variations across the ACIS field of view
(vignetting, chip gaps, etc) we generate effective exposure maps for
the observed XPS peak energy of 1.17 keV.\\

\section{Data reduction}

In order to maximize the detection efficiency we run the {\it
Celldetect}\/ algorithm in the 0.5--2.0 keV range\footnote{This
bandpass is selected to maximize the signal-to-noise ratio for sources
with AGN-like X-ray spectra}; only sources with detection significance
of at least $3\sigma$ above the local X-ray background are kept. For
each cluster we use the appropriate ACIS exposure map to identify and
remove spurious detections at the chip edges, leaving 910 point source
detections in the cluster fields and 509 detections in the control
fields.

Source count rates are converted to unabsorbed energy fluxes in the
0.5--2.0 keV energy band assuming the Galactic $n_{\rm H}$ value and a
power law with a spectral index of $\Gamma = 1.7$, consistent with the
observed stacked XPS spectrum\footnote{The details of the chosen
spectral model are not critical as, in this energy band, the observed
flux depends only weakly on the chosen power law slope
\citep{cappi}.}. Spatial variations of the detector characteristics
are taken into account in the conversion by computing Redistribution
Matrix Files ({\it RMF}) and Auxiliary Response Files ({\it ARF}) for
each source individually. Using a power-law model of the XPS $\log
N-\log S$ obtained from a maximum-likelihood fit to the
surface-density distribution of the XPS detected in 23 cluster
observation of duration 17.5--20.5 ks, the most common exposure time,
we compute instrumental flux limits of 95\% completeness for each
cluster scaled to individual exposure times.  We find the source lists
for all fields to be complete to a global flux limit of $1.25\times
10^{-14}$ erg s$^{-1}$ cm$^{-2}$, corresponding to an on-axis exposure
time of 10 ks. For point sources in cluster fields, we convert
unabsorbed X-ray fluxes to luminosities assuming the point sources to
be at the redshifts of the respective target cluster. Details of the
properties of the target clusters as well as of the XPS population
found in the corresponding ACIS-I fields will be presented in a
forthcoming paper (Ebeling et al., in preparation).

\section{Distribution of Point Sources}

\begin{figure}
\epsscale{1.2}
\plotone{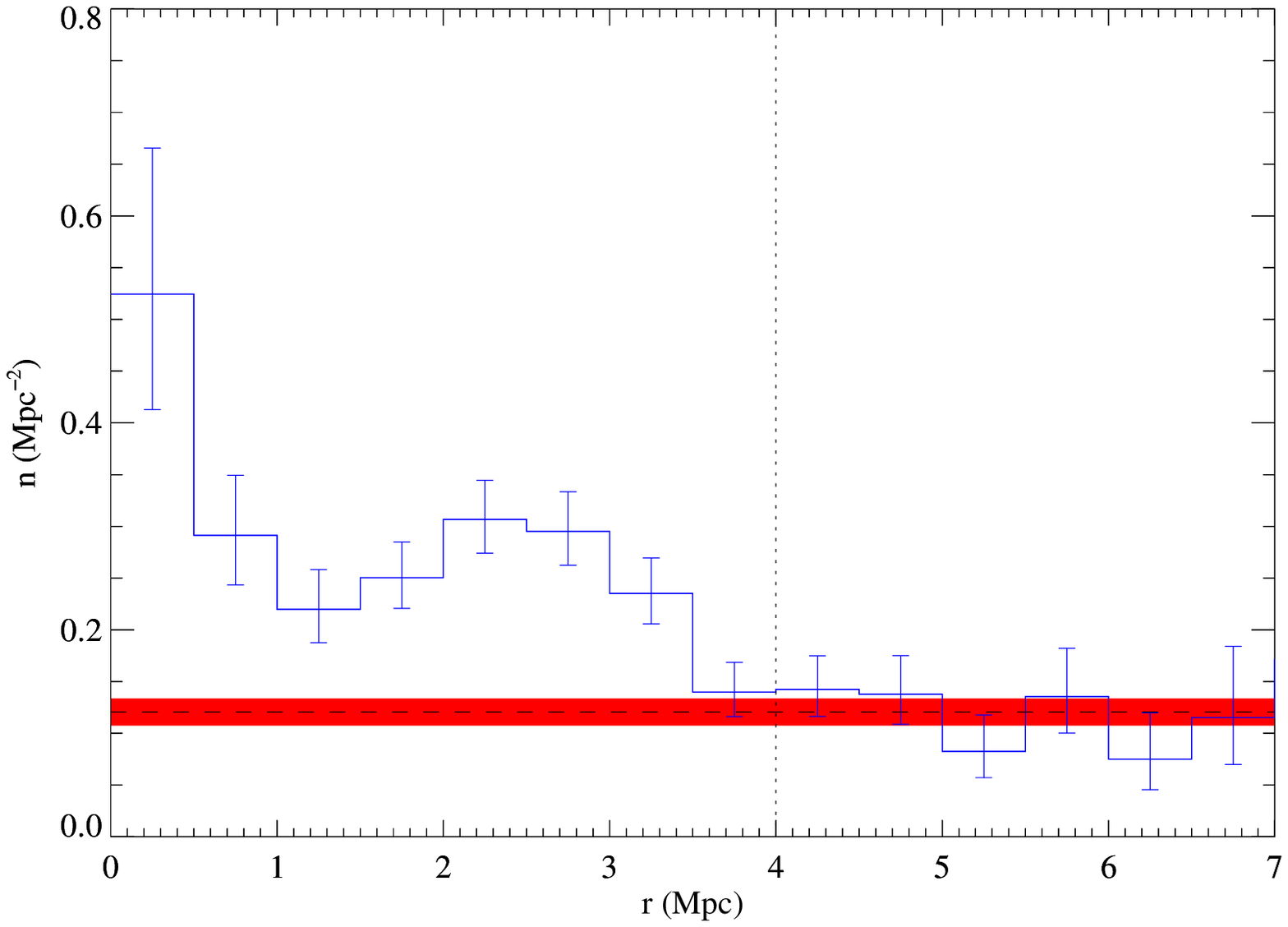}
\plotone{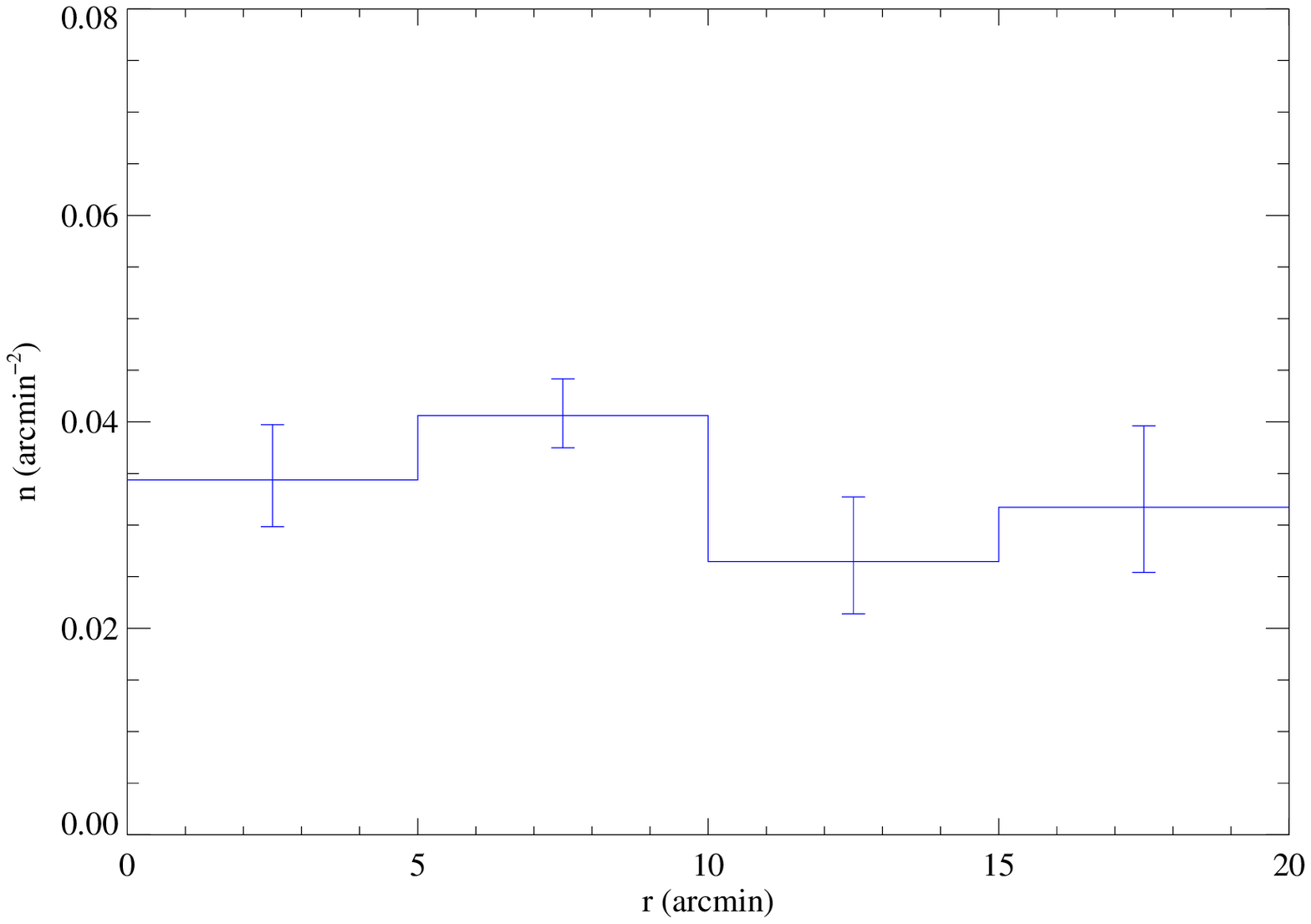}
\caption{\label{fig2} Radial profiles of the XPS surface density in 51
cluster fields (top) and 20 control fields (bottom). In the cluster
fields, the XPS density is constant beyond distances of about 3.5 Mpc
from the cluster center. The red bar marks this background level and
its $1\sigma$ error as derived from a fit to the data in the 4--7 Mpc
region. Within 3.5 Mpc a highly significant ($8.0\sigma$) excess is
observed in the cluster fields. By contrast the XPS surface density in
the control fields exhibits no excess. Both graphs probe similar
angular scales because, at the redshifts of our clusters, one Mpc
corresponds to 2.3 to 3.7 arcminutes.}
\end{figure}

We compute the metric distance from the cluster center for all XPS
detected in cluster fields, assuming again that the XPS are at the
cluster redshift. A radial profile of their surface density is then
constructed by binning the radial distances for the XPS from all 51
clusters, eliminating XPS fainter than the flux limit of the
respective observation (this leaves 508 XPS), and dividing the XPS
counts in each bin by the appropriate area. The result is shown in
Fig.~\ref{fig2} (top). The graph at the bottom of the same figure
shows a similar histogram for the 256 XPS detected above the
individual flux limits of the 20 control fields, using the angular
distances from the observation target position which corresponds to
the location on the detector where cluster centers fall in cluster
fields. For both panels of Fig.~\ref{fig2}, as well as for
Fig.~\ref{fig3}, the value of the surface density in each bin is
computed as $n = \sum_i N_i/\sum_i A_i$, with $N_i$ and $A_i$ being
the XPS number and area of the respective annulus for the $i^{\rm th}$
cluster; the shown $1\sigma$ error bars assume Poisson statistics using
the analytic approximations of \citet{poisson}.

The significance of the central XPS excess observed in the cluster
fields can be computed from the difference between the XPS surface
density observed in the inner 3.5 Mpc of the cluster fields and the
one observed in the outer 4--7 Mpc annulus, which we assume to be the
background XPS density. We find the excess to be significant at the
$8.0\sigma$ confidence level. No central excess is observed in the
radial XPS density profile for the control fields, ruling out a
systematic instrumental effect or processing error. Note that the
background per Mpc$^2$ implied by the control fields is higher than
the one measured in the outer regions of the cluster fields.  We
believe that this discrepancy reflects differences in the large-scale
environment between the cluster and control samples which we will
discuss in a future paper (Ebeling et al., in preparation).\\

\section{Correlation with cluster morphology}

In order to gain insight into the physical origin of the radial XPS
distribution shown in Fig.~\ref{fig2} (top) we crudely classify our 51
MACS clusters as relaxed or disturbed based on their X-ray morphology.
Using Chandra images adaptively smoothed with the ASMOOTH algorithm
\citep{asmooth}, we consider a cluster to be relaxed only if its X-ray
surface-brightness distribution exhibits the following
characteristics: a) on small scales, a pronounced central peak typical
of a cooling core, and b) on large scales, near-perfect spherical
symmetry. The combination of these criteria ensures that the clusters
thus selected have not undergone a major merger event in several Gyr
and can be considered virialized. For these systems the metric
distance used to construct the histogram shown in Fig.~\ref{fig2}
(top) corresponds to a well defined cluster-centric radius which
probes values of the basic cluster properties --- such as gas
temperature and density, or galaxy volocity dispersion --- that show
little variation with azimuthal angle.

\begin{figure}
\epsscale{1.2} \plotone{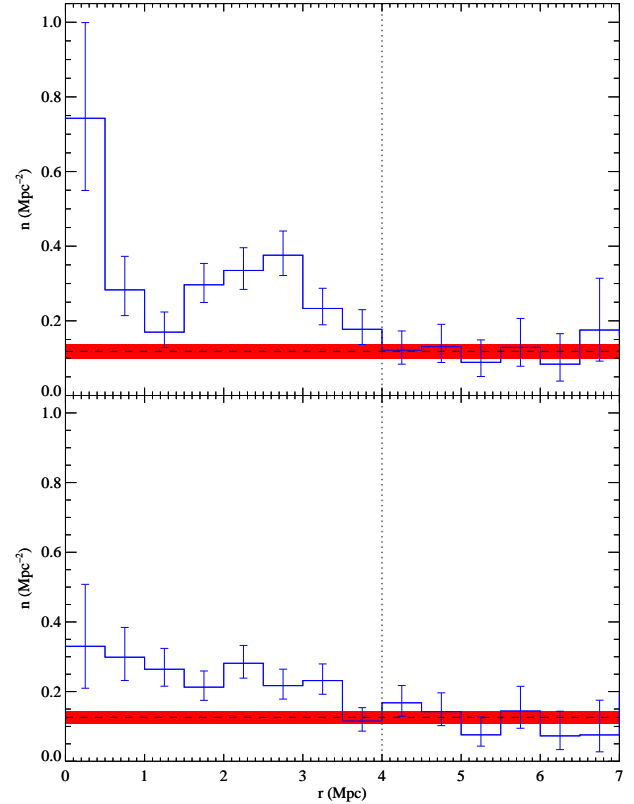}
\caption{\label{fig3} Radial profiles of the XPS surface density in
the fields of 24 relaxed clusters (top) and in the fields of 27
disturbed clusters (bottom). The profile for relaxed clusters features
a pronounced central peak at $r<0.5$ Mpc, a depletion region at
0.5--1.5 Mpc, and a broad secondary excess at approximately the virial
radius of 2--3 Mpc. By contrast, the XPS density profile for disturbed
clusters shows a smoothly distributed excess.}
\end{figure}

Our morphological classification splits our XPS list into two X-ray
flux limited subsamples, each containing 254 XPS, detected in the
fields of 24 relaxed or 27 disturbed systems,
respectively. Fig.~\ref{fig3} shows the radial XPS density profiles
for both subsets separately. Although both profiles exhibit a
prominent excess of XPS within approximately 3.5 Mpc of the cluster
center --- significant at 6.3 and 4.7$\sigma$ confidence for relaxed
and disturbed systems, respectively --- the two distributions are
markedly different (at $3.0\sigma$ signficance) with only the
spherically symmetric, relaxed clusters showing the double peak
apparent already in Fig.~\ref{fig2} (top).\\

\section{Correlation with cluster redshift}

We attempt to also test how the excess depends on cluster redshift.
In doing so, it is critical that the luminosities of the AGN probed at
different redshifts are comparable, i.e., completeness to a universal
{\it luminosity}\/ limit needs to be ensured. For a first, crude check
of any redshift dependence we divide our cluster sample into two
subsets, comprising 29 clusters at $z=0.3-0.45$ and 21 clusters at
$z=0.45-0.6$. Converting the flux limits of the different observations
into luminosities, we find that our XPS lists for the $z=0.3-0.6$
redshift range are 95\% complete to a global luminosity limit of
$1.1\times 10^{43}$ erg s$^{-1}$ (0.5--2.0 keV). Applying this
luminosity cut reduces our XPS sample to 87 and 119 sources in the
$z=0.3-0.45$ and $z=0.45-0.6$ cluster fields, respectively.  Although
the data are suggestive of an increase of the excess with redshift,
the difference between the two radial profiles is not significant
($1.5\sigma$) at the current sizes of our subsamples.
	
\section{Discussion}

\subsection{XPS nature}

The X-ray luminosities referred to in the previous section can be used
to characterize the nature of the point sources causing the observed
excess. As a result of the, in general, moderate exposure times of the
ACIS observations used in our study, we find all of the XPS in our
flux-complete sample to feature X-ray luminosities well in excess of
$3\times 10^{42}$ erg s$^{-1}$ (0.5--8.0 keV). Since this value
represents a tight upper limit to the X-ray luminosity of starburst
galaxies \citep{bauer} we conclude that the observed XPS excess can be
attributed to AGN at the cluster redshift.

The alternative explanation that the excess is caused by gravitational
lensing of background AGN (which would then have to be yet more X-ray
luminous) can be ruled out because of the large area over which the
excess is observed. The amplification provided by lensing can account
at best for a small fraction of the observed excess at $r<0.5$ Mpc,
and is entirely negligible at the angular distances corresponding to
the radius at which the secondary excess is observed \citep{smith}. In
addition, studies searching for an XPS excess in cluster fields
observed to much lower X-ray fluxes indicate that the faint end of the
$\log N-\log S$ distribution of the excess is inconsistent with the
one predicted for gravitationally lensed sources \citep{cappi}.
We therefore conclude that the excess is due to AGN in the cluster.

\subsection{XPS distribution}

We now discuss the possible origin of the spatial XPS distribution,
which is characterized by a central excess within 0.5 Mpc of the
cluster core, a depletion region around 1.5 Mpc, and a broad secondary
excess observed at about the virial radius of the relaxed
clusters. Although present in the fields of all clusters, the excess
at $r<0.5$ Mpc is much more pronounced for the relaxed systems, all of
which possess central cooling cores dominated by massive cD
galaxies. Closer inspection of the XPS distribution reveals that the
difference in this first radial bin between relaxed and disturbed
systems originates at yet smaller cluster-centric distances. We detect
six XPS within 250 kpc of the cD galaxies in virialized clusters and
none within the same distance of the cores of disturbed systems. Two
of these six XPS in the very heart of virialized clusters can be
unambiguously identified as AGN in the cD itself; for the other four
cases we propose that the observed X-ray emission is due to nuclear
activity triggered by close encounters between infalling cluster
galaxies and the cD. Extending this picture to the remainder of the
$r<0.5$ Mpc bin, we predict that the central excess is due to galaxy
mergers and tidal interactions involving the giant elliptical galaxies
dominating the centers of MACS clusters.

The AGN depletion zone in relaxed clusters at radii around 1.5 Mpc and
the following broad excess at roughly the virial radius can be
explained by a slightly different mechanism. The rise of the observed
XPS surface density at $2-3$ Mpc suggests increased AGN activity at
the cluster-field interface, which is easily explained by
merger-induced accretion onto massive black holes, favoured by the
increase in galaxy density compared to the field and the presence of
relatively gas-rich galaxies in this transition region.  Since
successful mergers require low collision velocities of typically less
than 300 km s$^{-1}$ \citep{binney} such mergers become extremely rare
closer to the cluster core where, for MACS clusters, the galaxy
velocity dispersion reaches and exceeds 1000 km s$^{-1}$ (Barrett et
al., in preparation). As a result, the AGN density is dramatically
reduced at intermediate radii (0.5--2 Mpc). In addition, activity
triggered at larger radii will cease as infalling galaxies approach
the cluster core since the timescale for the depletion of the
accretion disk is substantially shorter than the cluster crossing
time. The combination of these two effects ensures that the excess is
confined to the infall region where AGN activity is triggered.

We expect the same physical mechanisms to cause AGN activity in
disturbed clusters. However, since disturbed clusters lack well
defined cores as well as spherical symmetry, the excess is spread much
more evenly over the $0-3.5$ Mpc range, as observed (Fig.~\ref{fig3},
bottom).

\section{Conclusions}

We present results of the first study of the X-ray point source
content of a statistically well defined sample of galaxy clusters,
using Chandra/ACIS observations of 51 MACS clusters at $z=0.3-0.7$.
We detect an overall $8.0\sigma$ significant excess in the point
source density within 3.5 Mpc of cluster centers, which we argue is
caused by AGN in the cluster targets of these observations.  We also
present the first resolved radial profile of the excess and find it to
depend significantly on cluster morphology. Making use of the simple
geometry afforded by relaxed clusters we are able to identify two
distinct components to the excess, namely a central excess of AGN
which we believe to be due to galaxy interactions involving the giant
ellipticals near the cluster core, and a broad secondary excess at
about the virial radius.  We propose that the AGN activity causing the
secondary excess is triggered by galaxy mergers, which are most likely
to occur in the low-energy collisions favoured in the cluster-field
transition region. The lack of AGN at intermediate radii can be
explained by the low merger probability at high relative velocities
and the shortness of the accretion timescale compared to the cluster
crossing time.

Our results confirm those of past studies which have detected point
source excesses in cluster fields and provide a novel way to probe the
dynamics of AGN production by galaxy interactions in massive clusters.
Future work will focus on the optical counterparts of X-ray point
sources in selected MACS clusters to unambiguously identify redshifts
and X-ray emission processes, thus testing conclusively our hypothesis
that the excess is due to AGN cluster members.

\acknowledgments

The authors gratefully acknowledge financial support from NASA grant
NAG 5-8253 and SAO grant GO2-3168X (HE), and from the National Science
Foundation's Research Experiences for Undergraduates program (JTR). We
thank everybody at the Chandra X-Ray Center for their contribution to
a spectacularly successful satellite mission and specifically for
maintaining the CIAO software package.  Thanks also to Joshua Barnes
for a very informative discussion about the physics and consequences
of galaxy mergers, and to the referee, Stefano Ettori, for helpful
comments and criticism.


%% copy editing.  Individual instruments can be provided in parentheses,
%% after the keyword, but they will not be verified.

%Facilities: \facility{CXO(ACIS)}.

\end{document}